\documentclass[fleqn,usenatbib]{mnras}

\usepackage{newtxtext,newtxmath}

\usepackage[T1]{fontenc}

\DeclareRobustCommand{\VAN}[3]{#2}
\let\VANthebibliography\thebibliography
\def\thebibliography{\DeclareRobustCommand{\VAN}[3]{##3}\VANthebibliography}

\usepackage{graphicx}	
\usepackage{amsmath}	

\newcommand{\Msun}{\ensuremath{\mathrm{M}_\odot}}

\newcommand{\carcsec}{$\!\!\arcsec$}

\graphicspath{{./}{figures/}}

\title[GRSN rate]{
Constraint on the event rate of general relativistic instability supernovae from the early JWST deep field data
}

\author[T. J. Moriya et al.]{
Takashi J. Moriya,$^{1,2}$\thanks{E-mail: takashi.moriya@nao.ac.jp (TJM), hari@icrr.u-tokyo.ac.jp (YH)}
Yuichi Harikane,$^{3}$
and 
Akio K. Inoue$^{4,5}$
\\
$^{1}$National Astronomical Observatory of Japan, National Institutes of Natural Sciences, 2-21-1 Osawa, Mitaka, Tokyo 181-8588, Japan \\
$^{2}$School of Physics and Astronomy, Faculty of Science, Monash University, Clayton, Victoria 3800, Australia \\
$^{3}$Institute for Cosmic Ray Research, The University of Tokyo, 5-1-5 Kashiwanoha, Kashiwa, Chiba 277-8582, Japan \\
$^{4}$Department of Physics, Graduate School of Advanced Science and Engineering, Faculty of Science and Engineering, Waseda University, 3-4-1 Okubo, Shinjuku, \\
Tokyo 169-8555, Japan \\
$^{5}$Waseda Research Institute of Science and Engineering, Faculty of Science and Engineering, Waseda University, 3-4-1 Okubo, Shinjuku, Tokyo 169-8555, Japan
}

\date{Accepted 2023 September 21. Received 2023 September 9; in original form 2023 April 11}

\pubyear{2023}

\begin{document}
\label{firstpage}
\pagerange{\pageref{firstpage}--\pageref{lastpage}}
\maketitle

\begin{abstract}
General relativistic instability supernovae at $10\lesssim z \lesssim 15$ are predicted to be observed as red faint point sources, and they can be detected only in the reddest filters in JWST/NIRCam (\textit{F444W} and \textit{F356W}). They should be observed as persistent point sources with little flux variations for a couple of decades because of time dilation. We search for static point sources detected only in the \textit{F444W} filter or only in the \textit{F444W} and \textit{F356W} filters in the early JWST deep field data. No real point source of such kind is identified. Therefore, the general relativistic instability supernova rate at $10\lesssim z \lesssim 15$ is constrained to be less than $\sim 8\times 10^{-7}~\mathrm{Mpc^{-3}~yr^{-1}}$ for the first time. 
\end{abstract}




\section{Introduction}
General relativistic instability supernovae (GRSNe) are theoretically predicted explosions of supermassive stars (SMSs) having $10^4-10^5~\Msun$. Most SMSs are likely to collapse directly to black holes through general relativistic instability \citep[e.g.,][]{shibata2002}, but a fraction of SMSs are suggested to explode before collapsing to black holes \citep[e.g,][]{fuller1986,montero2012,chen2014,nagele2020,nagele2022,nagele2023}. They can have extremely high explosion energies of around $10^{55}~\mathrm{erg}$ \citep[e.g.,][]{chen2014} and thus they can become extremely luminous supernovae (SNe, e.g., \citealt{moriya2021}). Especially, \citet{moriya2021} show that GRSNe at high redshifts can become extremely red transients. They can be detected only in the \textit{F444W} filter if they appear at $13\lesssim z \lesssim 15$ and only in the \textit{F444W} and \textit{F356W} filters if they appear at $10\lesssim z \lesssim 13$ in JWST/NIRCam. GRSNe are predicted to have a plateau phase without a significant luminosity change lasting for about 2~years in the rest frame. Because of time dilation, the plateau phase can last for $20-30~\mathrm{years}$ if they appear at $10\lesssim z \lesssim 15$. Thus, high-redshift GRSNe are observed as persistent static red point sources for $20-30~\mathrm{years}$ after they reach the plateau phase.

Following the successful launch in December 2021, JWST has released early deep field imaging data from NIRCam. These data have already been used to identify record-breaking high-redshift galaxy candidates \citep[e.g.,][]{harikane2023,finkelstein2022,2023MNRAS.518.6011D,2022ApJ...940L..14N,2022ApJ...938L..15C,2022arXiv221206683B,2023MNRAS.519.1201A,2023ApJ...942L...9Y}. Some high-redshift galaxy candidates have also been suggested to be high-redshift SNe \citep{yan2023}, but further observations are required to confirm. In this work, we search for GRSN candidates in the early JWST deep field data, and provide a constraint on the GRSN rate at $10\lesssim z \lesssim 15$ for the first time. We adopt a $\Lambda$CDM cosmology with $H_0=70~\mathrm{km~s^{-1}~Mpc^{-1}}$, $\Omega_M = 0.3$, and $\Omega_\Lambda = 0.7$. All the magnitudes in this paper are in the AB system.

\begin{table*}
	\centering
	\caption{Early JWST deep field data and their redshift ranges for GRSN detection.}
	\label{tab:eventrate}
	\begin{tabular}{lccccccc} 
		\hline
		Field & Area & \textit{F444W} limit$^a$ & \textit{F356W} limit$^a$ & \textit{F277W} limit$^a$ & \multicolumn{2}{c}{GRSN detection redshift range$^b$} & Total volume$^c$ \\
		 & ($\mathrm{arcmin^2}$) & (mag) & (mag) & (mag) & \textit{F444W} and \textit{F356W} & \textit{F444W} only & ($\mathrm{Mpc^{3}}$) \\
		\hline
        SMACS J0723       & 11.0 & 29.6 & 29.9 & 29.8 &    $10.4\leq z\leq 13.4$ & $13.4\leq z\leq 16.0$ & $7.7\times 10^4$ \\
        GLASS             &  6.8 & 29.6 & 29.9 & 29.6 &    $10.2\leq z\leq 13.4$ & $13.4\leq z\leq 16.0$ & $5.0\times 10^4$ \\
        CEERS1            &  8.4 & 29.1 & 29.7 & 29.5 &    $10.1\leq z\leq 13.2$ & $13.2\leq z\leq 15.4$ & $5.8\times 10^4$ \\
        CEERS2            &  8.5 & 29.4 & 29.6 & 29.5 &    $10.1\leq z\leq 13.1$ & $13.1\leq z\leq 15.8$ & $6.2\times 10^4$ \\
        CEERS3            &  8.4 & 29.2 & 29.7 & 29.6 &    $10.2\leq z\leq 13.2$ & $13.2\leq z\leq 15.5$ & $5.7\times 10^4$ \\
        CEERS6            &  8.4 & 29.0 & 29.7 & 29.5 &    $10.1\leq z\leq 13.2$ & $13.2\leq z\leq 15.2$ & $5.6\times 10^4$ \\
        Stephan's Quintet & 37.2 & 28.6 & 28.9 & 28.8 & $\ \ 9.6\leq z\leq 12.4$ & $12.4\leq z\leq 14.7$ & $2.6\times 10^5$ \\
		\hline
  \multicolumn{8}{l}{$^a$ $5\sigma$ limiting magnitudes from \citet{harikane2023}. The limiting magnitudes are measured in a 0.\carcsec2-diameter circular aperture.} \\
  \multicolumn{8}{l}{$^b$ Based on the light-curve model in \citet{moriya2021}. The redshift range is determined by the $5\sigma$ limiting magnitudes. See Fig.~\ref{fig:expected}.} \\
  \multicolumn{8}{l}{$^c$ Comoving volumes within the GRSN detection redshift range including both "\textit{F444W} and \textit{F356W}" and "\textit{F444W} only".}
	\end{tabular}
\end{table*}

\section{Early JWST Deep Field Data}
We use the NIRCam images made in \citet{harikane2023}, and construct multi-band photometric catalogs in a similar manner to \citet{harikane2023}.
Here we briefly describe the data reduction and the catalog construction.
We refer to \citet{harikane2023} for more details.

We use the four JWST NIRCam data sets obtained in the early release observations (ERO) and early release science (ERS) programs, ERO SMACS J0723 and Stephan’s Quintet \citep{2022ApJ...936L..14P}, ERS Cosmic Evolution Early Release Science \citep{2022arXiv221105792F}, and ERS GLASS \citep{2022ApJ...935..110T}.
We reduced the raw data using the JWST pipeline version 1.6.3 development version (1.6.3.dev34+g6889f49, almost the same as 1.7.0), with the Calibration Reference Data System (CRDS) context file of {\tt jwst\_0995.pmap} released in October 2022, whose calibration values were derived using calibration observations of three different standard stars placed in all of the 10 NIRCam detectors.
In addition to the standard reduction by the pipeline, we conducted the ``wisp" subtraction by using a script provided by the NIRCam team\footnote{https://jwst-docs.stsci.edu/jwst-near-infrared-camera/nircam-features-and-caveats/nircam-claws-and-wisps}, the removal of the striping pattern by using a script provided in the CEERS team \citep{2022arXiv221102495B}, the sky background subtraction using {\sc SExtractor} \citep[version 2.5.0;][]{1996A&AS..117..393B}.
In the reduced images, the limiting magnitudes were measured with $0.\carcsec1$, $0.\carcsec2$, and $0.\carcsec3$-diameter circular apertures by randomly placing apertures in sky areas using Python packages {\sc Astropy/photutils}.
Sky areas were defined as pixels without objects detected by {\sc SExtractor}.
The effective area and limiting magnitudes are summarized in Table \ref{tab:eventrate}.

We construct multi-band source catalogs from the JWST data using the \textit{F444W} band as the detection band.
Source detection and photometry is performed with {\sc SExtractor} \citep[][]{1996A&AS..117..393B}.
We run {\sc SExtractor} in the dual-image mode for each image with its detection image (\textit{F444W}).
The total magnitudes are estimated with {\tt MAG\_AUTO} in {\sc SExtractor}.
Finally, we correct for the galactic extinction using \citet{1998ApJ...500..525S} and \citet{2011ApJ...737..103S} and make final photometric catalogs.

\section{Searching for GRSN events}
\begin{figure}
	\includegraphics[width=\columnwidth]{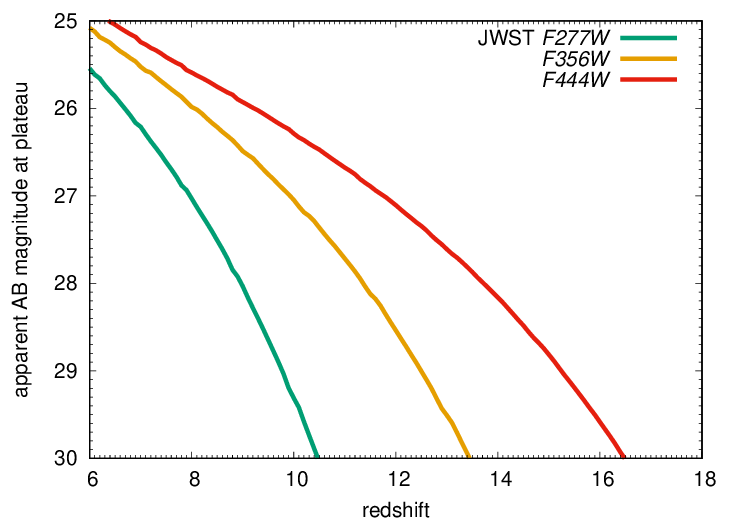}
    \caption{
    Expected brightness during the light-curve plateau of the GRSN predicted by \citet{moriya2021}.
    }
    \label{fig:expected}
\end{figure}

Figure~\ref{fig:expected} presents expected magnitudes of the GRSN model at its light-curve plateau phase presented by \citet{moriya2021}. Table~\ref{tab:eventrate} presents the redshift ranges in which we expect to observe GRSNe in the \textit{F444W} and \textit{F356W} filters with the $5\sigma$ limiting magnitudes in each JWST deep field based on the GRSN light-curve model. The redshift ranges depend on the field because each field has different $5\sigma$ limiting magnitudes. First, we searched for static point sources that are not detected in the \textit{F277W} and \textit{F356W} filters with the $5\sigma$ significance, but are detected in the \textit{F444W} filter with more than the $5\sigma$ significance. 
Source detections are evaluated  based on fluxes measured in the $0.\carcsec1$ and/or $0.\carcsec2$-diameter circular apertures in the original (not PSF-matched) images.
We searched for static point sources that are fainter than 26.0~mag, because the GRSN model predicts that it becomes fainter than 26.0~mag at $z\gtrsim 10$ (Fig.~\ref{fig:expected}). The definition of a point source in this paper is sources with ${\tt CLASS\_STAR} > 0.9$. We found 60 objects matching these criteria. We checked them by eyes and we identified them as either cosmic rays or artifacts because of their shapes and extensions. 

We also searched for faint ($> 26.0~\mathrm{mag}$) static point sources that are detected with more than $5\sigma$ significance in both \textit{F356W} and \textit{F444W} filters, but not detected in the \textit{F277W} filter with the $5\sigma$ significance. We only found three candidate objects. The significant reduction in the number of candidates from the single detection in the \textit{F444W} filter indicates that most candidates observed only in the \textit{F444W} filter are likely artificial, because a chance to detect random artificial objects at the same location in both the \textit{F444W} and \textit{F356W} images is presumed to be low. Through the visual inspection, we identified them as artifacts because of their shapes and extensions. The three candidates are found to be either near the edge of a detector or blended with another object. These facts may have caused artifacts.

\begin{figure}
	\includegraphics[width=\columnwidth]{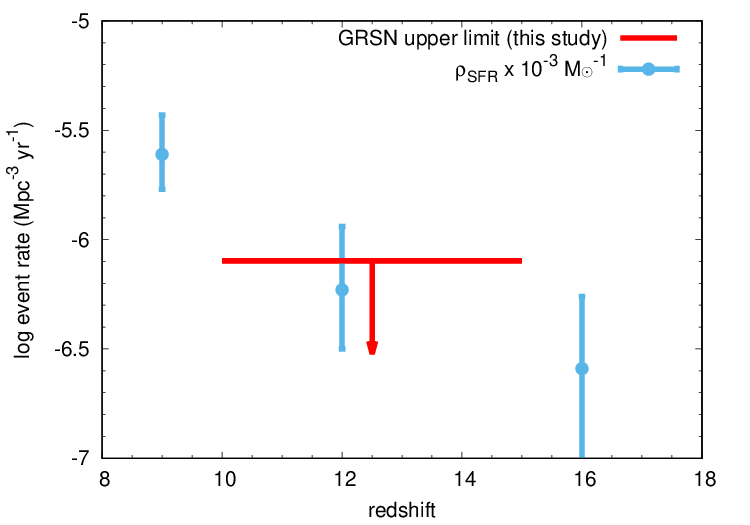}
    \caption{
    Upper limit for the GRSN event rate obtained by this work ($\lesssim 8\times 10^{-7}~\mathrm{Mpc^{-3}~yr^{-1}}$ at $10\lesssim z\lesssim 15$). The current GRSN event rate constraint and the estimated SFR density $\rho_\mathrm{SFR}$ at $z\sim 9-16$ by \citet{harikane2023} indicate that the GRSN event rate should be less than around $\rho_\mathrm{SFR}\times 10^{-3}~\Msun^{-1}$.
    }
    \label{fig:rate}
\end{figure}

\section{Constraint on GRSN event rate}\label{sec:rate}
As described in the previous section, we did not find any real static point sources fainter than 26.0~mag that are detected only in the \textit{F444W} filter. We also did not find any real static point sources fainter than 26.0~mag that are detected only in the \textit{F444W} and \textit{F356W} filters. These facts show that there has been no GRSN event in the survey fields in about $2~\mathrm{years}$ in the rest frame ($\sim 20-30~\mathrm{years}$ in the observer frame) in the redshift ranges shown in Table~\ref{tab:eventrate}. The exact redshift ranges that each field can constrain are slightly different because of the different limiting magnitudes, but they are roughly at $10\lesssim z \lesssim 15$. The total comoving volume of the survey fields is $6.2\times 10^5~\mathrm{Mpc^3}$. The upper limit for the GRSN event rate was estimated by assuming that the number of GRSNe in this volume in the last 2~years is less than one. In this way, the GRSN event rate at $10\lesssim z\lesssim 15$ is estimated to be less than about $8\times 10^{-7}~\mathrm{Mpc^{-3}~yr^{-1}}$ in the comoving frame.

Cosmic star-formation rate (SFR) density $\rho_\mathrm{SFR}$ at $z\sim 9-16$ has been constrained by \citet{harikane2023}. If a fraction $f_\mathrm{GRSN}$ of SMSs having $10^4-10^5~\Msun$ explodes as GRSNe, the GRSN event rate $R_\mathrm{GRSN}$ can be expressed as
\begin{equation}
   R_\mathrm{GRSN}= \rho_\mathrm{SFR}f_\mathrm{GRSN}\Psi(\Gamma), 
\end{equation}
where
\begin{equation}
    \Psi(\Gamma) \equiv \frac{\int_{10^4~\Msun}^{10^5~\Msun}\psi (M,\Gamma)dM}{\int_{0.1~\Msun}^{10^5~\Msun}M\psi (M,\Gamma)dM},
\end{equation}
and $\psi(M,\Gamma) \propto M^{-\Gamma}$ is the initial mass function (IMF). We assume that stars between 0.1~\Msun\ and $10^5~\Msun$ are formed at the redshift ranges we are interested in. Figure~\ref{fig:rate} shows that our GRSN event rate constraint provide the following constraint at $10\lesssim z\lesssim 15$,
\begin{equation}
    f_\mathrm{GRSN}\Psi(\Gamma) \lesssim 10^{-3}~\Msun^{-1}. \label{eq:const}
\end{equation}

The Salpeter IMF ($\Gamma=2.35$) gives $\Psi(2.35)=4\times 10^{-7}~\Msun^{-1}$ and the flat IMF ($\Gamma=0$) gives $\Psi(0)=2\times 10^{-5}~\Msun^{-1}$. Thus, with the current constraint (Eq.~\ref{eq:const}), the fractions of SMSs exploding as GRSNe are constrained to be $f_\mathrm{GRSN}\lesssim 3000$ for the Salpeter IMF and $f_\mathrm{GRSN}\lesssim 50$ for the flat IMF. Because these constraints are well above $f_\mathrm{GRSN}= 1$, our upper limit is still satisfied even if all SMSs explode as GRSNe. Indeed, SMS formation rates at $10\lesssim z \lesssim 15$ are predicted to be $10^{-8}-10^{-12}~\mathrm{Mpc^{-3}~yr^{-1}}$ \citep[e.g.,][]{agarwal2012,chon2016,habouzit2016,dunn2018,chiaki2023}, which is lower than our upper limit estimates of $\sim 8\times 10^{-7}~\mathrm{Mpc^{-3}~yr^{-1}}$. A further, more meaningful constraint on the GRSN rate and its fraction can be obtained as the JWST Deep Field areas become wider.

\section{Summary}\label{sec:conclusions}
The GRSN event rate at $10\lesssim z\lesssim 15$ has been estimated to be $\lesssim 8\times 10^{-7}~\mathrm{Mpc^{-3}~yr^{-1}}$ based on the fact that no GRSN candidates are discovered in the early JWST deep field data. The estimated upper limit of the GRSN event rate ($\sim 8\times 10^{-7}~\mathrm{Mpc^{-3}~yr^{-1}}$) is still too high to constrain the formation rate of SMSs and the fraction of SMSs that explode as GRSNe. As the area of the JWST deep field increases, we will be able to further constrain the GRSN rate. We encourage to search for faint point sources that are detected only in \textit{F444W} or only in both \textit{F444W} and \textit{F356W} in the future JWST deep images. They will provide us critical information on the formation rate and fate of SMSs that can be compared to theoretical predictions of the first star formation.

\section*{Acknowledgements}
TJM thanks Sunmyon Chon and Gen Chiaki for useful discussions.
This work is supported by the Grants-in-Aid for Scientific Research of the Japan Society for the Promotion of Science (JP20H00174, JP21K13966, JP21H04997, JP21K13953, JP23H00131).
This work is based on observations made with the NASA/ESA/CSA James Webb Space Telescope. The data were obtained from the Mikulski Archive for Space Telescopes at the Space Telescope Science Institute, which is operated by the Association of Universities for Research in Astronomy, Inc., under NASA contract NAS 5-03127 for JWST. These observations are associated with programs 2732, 2736, 1324, and 1345.
The authors acknowledge the
ERO, GLASS, and CEERS teams
led by
Klaus M. Pontoppidan,
Tommaso Treu, and
Steven L. Finkelstein, respectively,
for developing their observing programs with a zero-exclusive-access period.

\section*{Data Availability}
The data underlying this article will be shared on reasonable request to the corresponding author.



\bibliographystyle{mnras}
\bibliography{mnras} 






\bsp	
\label{lastpage}
\end{document}